\documentstyle[prl,aps,epsf]{revtex}
\draft
\begin{document}

\title{ {\tt {\small \begin{flushright}
hep-ph/0408103\\
August 2004
\end{flushright} } }
Resonant {\boldmath $\tau$}-Leptogenesis with Observable Lepton Number
Violation }

\author{Apostolos Pilaftsis}

\address{School of Physics and Astronomy, University of Manchester,\\ 
Manchester M13 9PL, United Kingdom}

\maketitle

\begin{abstract}

\noindent
We consider a minimal extension of the Standard Model with one singlet
neutrino per generation that  can realize resonant leptogenesis at the
electroweak scale. In particular, the baryon asymmetry in the Universe
can be created by  lepton-to-baryon conversion of an individual lepton
number, for  example that of the $\tau$-lepton.   The current neutrino
data can  be explained by  a simple CP-violating Yukawa  texture.  The
model has several testable phenomenological implications.  It contains
heavy Majorana neutrinos at the electroweak scale, which can be probed
at    $e^+e^-$    linear   colliders,    and    predicts   $e$-    and
$\mu$-lepton-number-violating  processes,   such  as  $0\nu\beta\beta$
decay,  $\mu \to  e\gamma$ and  $\mu$-$e$ conversion  in  nuclei, with
rates that are within reach of experimental sensitivity.
\medskip

\end{abstract}

\bigskip

The origin of our matter--antimatter-asymmetric Universe is one of the
central  themes in particle  cosmology.  In  the  light of the  recent
high-precision observation   of the baryon-to-photon  ratio  of number
densities, $\eta_B \approx  6.1 \times  10^{-10}$~\cite{WMAP}, finding
a~laboratory testable solution    to this problem  becomes   even more
motivating. A consensus has  been reached that  a possible solution to
the problem  of the baryon   asymmetry in the Universe~(BAU)  requires
physics beyond  the   Standard  Model~(SM).    In this  context,    an
interesting suggestion has been  that  neutrinos, which  are  strictly
massless in the SM, may acquire their observed  tiny mass naturally by
the  presence   of superheavy  partners   through the so-called seesaw
mechanism~\cite{seesaw}.   These  superheavy  neutrinos have  Majorana
masses  that  violate the lepton   number~($L$) by  two  units.  Their
out-of-equilibrium  decay   in  an  expanding Universe   may initially
produce   a  leptonic asymmetry, which    is  then converted into  the
observed   BAU~\cite{FY}   by equilibrated $(B+L)$-violating sphaleron
interactions~\cite{KRS}.

In ordinary  seesaw models embedded in grand  unified theories (GUTs),
the natural mass scale of  the heavy Majorana neutrinos is expected to
be  of order  the  GUT scale  $10^{16}$~GeV.   However, the  reheating
temperature $T_{\rm  reh}$ in these  theories is generically  of order
$10^9$~GeV, thus requiring for one of the heavy neutrinos, e.g.~$N_1$,
to be  unnaturally light below $T_{\rm  reh}$, so as  to be abundantly
produced  in  the  early  Universe.   On the  other  hand,  successful
leptogenesis and  compatibility with the existing neutrino  data put a
lower   bound  on   the  $N_1$-mass:   $m_{N_1}  \stackrel{>}{{}_\sim}
10^9$~GeV~\cite{DI}.  To avoid this  narrow window of viability of the
model around $10^9$~GeV, one needs  to assume that the second heaviest
neutrino  $N_2$ is  as light  as $N_1$~\cite{CT},  an  assumption that
makes this thermal GUT leptogenesis scenario even more unnatural.

A solution that avoids the  aforementioned $T_{\rm reh}$ problem is to
have   low-scale    thermal   leptogenesis~\cite{APRD,APreview}.    To
accomplish  this,  one  may   exploit  the  fact  that  heavy-neutrino
self-energy  effects~\cite{LS,Paschos} on  the leptonic  asymmetry get
resonantly enhanced, even up to order~1, when a pair of heavy Majorana
neutrinos has a mass difference comparable to the heavy neutrino decay
widths~\cite{APRD}.  In  this case, the  scale of leptogenesis  can be
lowered  to the TeV  range~\cite{APRD,APreview} in  complete agreement
with the current neutrino  data~\cite{PU}.  Even though the discussion
will be  focused on  a minimal non-supersymmetric  3-generation model,
the   ideas  presented  here   could  be   extended  to   unified  and
supersymmetric theories as well~\cite{GCBetal,HMW}.

In this  Letter we study  a potentially important variant  of resonant
leptogenesis,  where a  given individual  lepton number  is resonantly
produced by out-of-equilibrium decays of heavy Majorana neutrinos of a
particular family type.  For the  case of the $\tau$-lepton number, we
call  this   mechanism  {\em  resonant   $\tau$-leptogenesis}.   Since
sphalerons  preserve the  individual quantum  numbers  $\frac{1}{3}B -
L_{e,\mu,\tau}$~\cite{HT} in addition to the $B - L$ number, an excess
in  $L_\tau$ will  be converted  into the  observed BAU,  provided the
possible $L_\tau$-violating  reactions are  not strong enough  to wash
out such an excess.  Moreover, a chemical potential analysis~\cite{HT}
shows that  the generated baryon  asymmetry is $B =  - \frac{28}{51}\,
L_\tau$ at  temperatures $T$  above the electroweak  phase transition,
i.e.~for $T  \stackrel{>}{{}_\sim} T_c \approx  150$--200~GeV.  Hence,
generating  the BAU from  an individual  lepton-number excess  is very
crucial for the  resonant $\tau$-leptogenesis scenario presented below
to  have  phenomenologically  testable  signatures  of  lepton  number
violation.

The model under discussion  is the SM  symmetrically extended with one
singlet neutrino $\nu_R$ per family.   The leptonic flavour  structure
of the Yukawa and Majorana sectors of such a model may be described by
the Lagrangian
\begin{equation}
  \label{Lym}
-\, {\cal L}_{\rm Y,M}\ =\ 
\frac{1}{2}\, (\bar{\nu}_{iR})^C (M_S)_{ij}\, \nu_{jR}\ +\ 
\hat{h}^l_{ii}\, \bar{L}_i\,\Phi\, l_{iR} \ +\
h^{{\nu_R}}_{ij}\,
\bar{L}_i\, \tilde{\Phi}\, \nu_{jR} \ +\ {\rm h.c.}\,,
\end{equation}
where $i,j = 1,2,3$ and $\tilde{\Phi}$ is the isospin conjugate of the
Higgs doublet $\Phi$.     We  define the individual   lepton   numbers
in~(\ref{Lym})  in the  would-be charged-lepton mass  basis, where the
charged-lepton Yukawa matrix  $\hat{h}^l$ is diagonal.   Note that all
SM  reactions, including  those that involve  the $e$-Yukawa coupling,
will      be   in   thermal    equilibrium     for   temperatures   $T
\stackrel{<}{{}_\sim} 10$~TeV~\cite{CKO,BCST}, relevant   to low-scale
leptogenesis models.  The calculation of the BAU  will be performed in
the heavy   neutrino mass  basis,   where the  3 heavy  neutrino  mass
eigenstates  are denoted by  $N_{1,2,3}$.   Such a selection could  be
justified   from  arguments     based on  decoherentional   in-thermal
equilibrium dynamics, and by   the fact that heavy  Majorana  neutrino
decays      are thermally   blocked    already   at  temperatures   $T
\stackrel{>}{{}_\sim} 3 m_{N_i}$~\cite{GNRRS}.

We  now      present    a    generic     scenario     for     resonant
$\tau$-leptogenesis. The neutrino  Yukawa sector of this  scenario has
the following maximally CP-violating form:
\begin{equation}
 \label{hmatrix}
h^{\nu_R} \ =\ \left(\! \begin{array}{ccc}
\varepsilon_e    & a\, e^{-i\pi/4}  & a\,e^{i\pi/4} \\
\varepsilon_\mu  & b\, e^{-i\pi/4}  & b\,e^{i\pi/4} \\
\varepsilon_\tau & c\, e^{-i\pi/4}  & c\,e^{i\pi/4} \end{array} \!\right)\, .
\end{equation}
For  nearly  degenerate $N_i$  masses   in the  range $m_{N_i} \approx
0.5$--1~TeV, the moduli of  the  complex parameters  $a,b$ have to  be
smaller  than about~$10^{-2}$  for  phenomenological  reasons  to   be
discussed below.   On  the other  hand, the requirement  to protect an
excess in  $L_\tau$  from wash-out   effects  leads to the  relatively
stronger constraint $|c| \stackrel{<}{{}_\sim} 10^{-4}$.  Furthermore,
the parameters $\varepsilon_l$, with $l = e,\mu,\tau$, are taken to be
small   perturbations  of  the   order  of  the  $e$-Yukawa  coupling,
i.e.~$|\varepsilon_l| \sim 10^{-6}$.   Here, we should  stress that at
least  3 singlet    heavy    neutrinos  are  needed   to   obtain    a
phenomenologically relevant model.

It  is now  important  to  notice that  for  exactly degenerate  heavy
Majorana neutrinos $M_S = m_N\,{\bf 1}_3$ and $\varepsilon_l = 0$, the
light-neutrino mass matrix vanishes identically, i.e.~$(m^\nu)_{ll'} =
-\, v^2  h^{\nu_R}_{li} (M^{-1}_S)_{ij}\, h^{\nu_R}_{l'j}  = 0$, where
$v \approx  175$~GeV is  the usual SM  vacuum expectation  value.  The
resulting vanishing of the light neutrino masses will be an all-orders
result, protected  by a U(1)$_l$ leptonic symmetry,  where each lepton
doublet   couples  to   the   linear  combination:   $(\nu_{2R}  +   i
\nu_{3R})$~\cite{BGL}.    If  the  symmetry-breaking   parameters  are
$|\varepsilon_l| \sim  10^{-6}$ and $(\Delta M_S)_{ij}  = (M_S)_{ij} -
m_N\,\delta_{ij}          \stackrel{<}{{}_\sim}          10^{-7}\times
m_N$\footnote[1]{We  shall not address  here in  detail the  origin of
these small breaking parameters,  but they could result from different
sources, e.g.~the Froggatt--Nielsen mechanism~\cite{PU,FN}, Planck- or
GUT-scale lepton-number breaking~\cite{APRD,MV}, etc.}, the entries of
$m^\nu$ fall within the observed region of less than $\sim 0.1$~eV for
$m_N  \sim  1$~TeV.  At  this  point,  we  should emphasize  that  our
scenario  is   radiatively  stable.   For   a  TeV-scale  leptogenesis
scenario,  renormalization-group running effects  on $m^\nu$  are very
small~\cite{Antusch}.    In    addition,   there   are    Higgs-   and
$Z$-boson-mediated   threshold   effects   $\delta   m^\nu$   of   the
form~\cite{AZPC}:
\begin{equation}
  \label{mnumass}
(\delta m^\nu)_{ll'} \ \sim\ 10^{-3}\, 
\frac{v^2}{m^2_N}\:
h^{\nu_R}_{li}\,(\Delta M_S)_{ij}\, h^{\nu_R}_{l'j}\; .
\end{equation}
Given the constraints  on the Yukawa  parameters  discussed above, one
may estimate  that the  finite  radiative effects  $\delta m^\nu$ stay
well  below  0.01~eV for  $(\Delta M_S)_{ij}/m_N \stackrel{<}{{}_\sim}
10^{-7}$  and  $|a|,\ |b| \stackrel{<}{{}_\sim} 10^{-2}$.   Hence, the
perturbation parameters    $\varepsilon_l$ and  $(\Delta    M_S)_{ij}$
provide sufficient   freedom to describe   the existing neutrino data.
For our resonant $\tau$-leptogenesis  scenario, the favoured  solution
is  an inverted    hierarchical neutrino  mass   spectrum  with  large
$\nu_e\nu_\mu$ and  $\nu_\mu\nu_\tau$ mixings~\cite{PDG}.  An explicit
example  for    a scenario    with    $m_N   = 500$~GeV    is    given
in~\cite{Example}.

It is instructive to give  an order-of-magnitude  estimate of the  BAU
generated by resonant $\tau$-leptogenesis.  In  such a model, only the
heavy Majorana neutrino  $N_1$ will decay   relatively out of  thermal
equilibrium.   Instead, $N_2$     and  $N_3$ will  decay  in   thermal
equilibrium   predominantly into $e$   and   $\mu$ leptons.  To  avoid
erasure of a  potential $L_\tau$ excess, the  decay rates of $N_2$ and
$N_3$ to $\tau$-leptons should  be rather suppressed.  To be specific,
in this framework the predicted BAU is expected to be
\begin{equation}
  \label{BAU}
\eta_B\ \sim\ -\,10^{-2}\, \frac{\delta^\tau_{N_1}}{K_{N_1}}\ 
\frac{\Gamma (N_1 \to L_\tau\Phi)}{\Gamma (N_{2,3}\to L_\tau\Phi)}\ 
\sim\ -\,10^{-2}\, \frac{\delta^\tau_{N_1}}{K_{N_1}}\ 
\frac{\varepsilon^2_\tau}{c^2}\ ,
\end{equation}
where   $\delta^\tau_{N_1}$, computed analytically  in~\cite{PU} for a
3-generation model, is   the $\tau$-lepton  asymmetry  and $K_{N_1}  =
\Gamma_{N_1}/H(z\!=\!1)$ is   an  out-of-equilibrium  measure of   the
$N_1$-decay rate $\Gamma_{N_1}$ with respect to  the Hubble rate $H(z)
\approx 17\, m^2_{N_1}/(z^2\, M_{\rm Planck} )$, with $z = m_{N_1}/T$.
For  $\varepsilon_l \sim 10^{-6}$, it is  $K_{N_1} \sim 10$.  The size
of   $\eta_B$ is   determined    by the  key  parameters:   $K_{N_1}$,
$\delta^\tau_{N_1}$   and $\varepsilon_\tau/c$.   To account  for  the
observed BAU,  one would need, e.g.~$|\delta^\tau_{N_1}| \sim 10^{-5}$
and $\varepsilon_\tau/c  \sim 10^{-1}$.  Instead,  the parameters  $a$
and  $b$ could be as large  as $\sim 10^{-2}$, potentially giving rise
to observable effects   of  lepton-number violation  at colliders  and
laboratory experiments (see discussion below).

We  now perform a  simplified numerical  analysis of  the BAU  in this
scenario  of resonant $\tau$-leptogenesis.   Detailed results  of this
study  are given  in~\cite{PU1}.  Since  the $N_{2,3}$  heavy Majorana
neutrinos will  be in thermal  equilibrium, their contribution  to the
$L_\tau$ asymmetry will be vanishingly small. A conservative numerical
estimate may be obtained by solving the Boltzmann equation~(BE):
\begin{eqnarray}
  \label{BE}
\frac{d\eta_{L_\tau}}{dz}\ &=&\ -\,\frac{z}{H(z=1)\,n_\gamma}\, 
\bigg[\, \delta^\tau_{N_1}\, 
\bigg( 1\: -\: \frac{\eta_{N_1}}{\eta^{\rm eq}_{N_1}}\,\bigg)\,
\gamma^{N_1}_{L_\tau\Phi}\ +\ \frac{\eta_{L_\tau}}{4}\, \bigg(\,
\sum_{i=1}^3\,\gamma^{N_i}_{L_\tau\Phi}\:
+\: \gamma^S_{\not L_\tau}\, \bigg)\,\bigg]\, ,
\end{eqnarray}
where  $n_\gamma$ is the photon number   density, and $\eta_{N_1}$ and
$\eta_{L_\tau}$           are        the   $N_1$-number-to-photon  and
$L_\tau$-number-to-photon    ratios of number densities, respectively.
We follow the    conventions  of~\cite{PU} for the    collision  terms
$\gamma^{N_i}_{L_l\Phi}$  related to the  decays and inverse decays of
the  $N_{1,2,3}$  neutrinos into   $L_{e,\mu,\tau}$ and   $\Phi$.  The
collision      term    $\gamma^S_{\not    L_\tau}$     describes   the
$L_\tau$-violating  $2\leftrightarrow 2$ scatterings: (i)~$L_\tau \Phi
\leftrightarrow L^C_{e,\mu,\tau}  \Phi^\dagger$    and  $L_\tau   \Phi
\leftrightarrow    L_{e,\mu}   \Phi$;  (ii)~$L_\tau   L^C_{e,\mu,\tau}
\leftrightarrow  \Phi\Phi$  and    $L_\tau  L_{e,\mu}  \leftrightarrow
\Phi\Phi^\dagger$.  The collision terms  for the reactions~(i)  can be
shown          to           be      always      smaller           than
$\sum_{i=1}^3\,\gamma^{N_i}_{L_\tau\Phi}$  for the temperature  domain
$z\stackrel{<}{{}_\sim}  10$   of    interest, while  those   for  the
reactions~(ii) are suppressed by  factors proportional to  $(m_{N_i} -
m_{N_j})/T$, $(m^\nu)^2_{ij}/v^2$  and $(a+b)^2 \sim 10^{-4}$, and can
therefore  be  neglected.  To  obtain a  conservative estimate, we set
$\gamma^S_{\not L_\tau} = \sum_{i=1}^3\,\gamma^{N_i}_{L_\tau\Phi}$.

\begin{center}
\begin{figure}[b]
   \leavevmode
   \epsfxsize=8.cm
   \epsfysize=7.cm
    \epsffile[0 0 700 700]{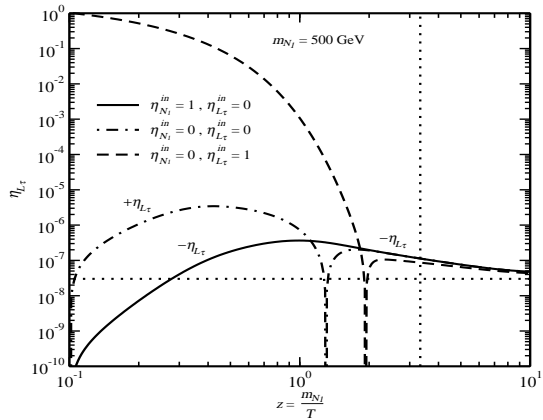} 
\medskip
\caption{Numerical  values for  $\eta_{L_\tau}$  versus $z=m_{N_1}/T$,
for different initial  conditions for $\eta_{N_1}$ and
$\eta_{L_\tau}$, and for the model parameters given in~[24].
The horizontal dotted line at $\eta_{L_\tau} \approx -3\times 10^{-8}$
indicates  the value  needed to  obtain  the observed  BAU, while  the
vertical dotted line corresponds to $T = T_c = 150$~GeV.}\label{f1}
\end{figure}
\end{center}

Figure~\ref{f1}   shows  numerical   estimates  of $\eta_{L_\tau}$  as
functions  of  $z = m_{N_1}/T$,  for  the resonant $\tau$-leptogenesis
scenario  given  in~\cite{Example}.    In   this  scenario,    it   is
$\delta^\tau_{N_1} \approx -  0.3\times 10^{-5}$ and $K_{N_1}  \approx
15$.  We find that $\eta_{L_\tau} (T_c)$ is independent of the initial
values $\eta^{\rm in}_{N_1}$ and  $\eta^{\rm in}_{L_\tau}$.  Thus, the
heavy  neutrino mass  $m_{N_1}$  could in principle be  as  low as the
critical   temperature   $T_c$,  namely   at   the electroweak  scale.
Employing    the lepton-to-baryon     conversion formula  for    $T\ll
T_c$~\cite{PU}, $\eta_B   \approx - \frac{1}{54}\,  \eta_{L_\tau}$, we
find  that   our   numerical    results are  compatible      with  the
estimate~in~(\ref{BAU}).  Instead, we have  carefully checked that the
usual formalism  of  BEs does not  properly  incorporate single lepton
flavour effects and leads to an erroneous  result for the BAU which is
a factor $\eta_L/\eta_{L_\tau} \sim (\delta_{N_1}/\delta^\tau_{N_1})\:
[|c|^2/(|a|^2 +|b|^2)] \sim 10^{-8}$~\cite{deltaN} too small!

\vspace{-0.5cm}

Our model of resonant $\tau$-leptogenesis has several phenomenological
consequences.  Since  the  model realizes an   inverted light-neutrino
mass spectrum~\cite{Example,Klapdor/Sarkar}, it  leads to  a  sizeable
neutrinoless  double beta ($0\nu\beta\beta$)   decay with an effective
neutrino  mass $|(m^\nu   )_{ee}|  \approx 0.013$~eV, which could   be
tested  in  future  $0\nu\beta\beta$    experiments.   The  model also
predicts $\mu \to e\gamma$ with a branching ratio $B(\mu \to e\gamma )
\approx    6   \cdot  10^{-4}\times    (a^2b^2    v^4)/m^4_N$  in  the
heavy-neutrino limit~\cite{CL}.   Confronting this prediction with the
experimental     limit    $B^{\rm   exp}     (\mu    \to   e\gamma   )
\stackrel{<}{{}_\sim} 1.2\times  10^{-11}$~\cite{PDG},  the  resulting
constraint  is    $(ab   v^2)/m^2_N  \stackrel{<}{{}_\sim}   1.4\times
10^{-4}$.  For electroweak-scale heavy neutrinos and $a,b \sim 3\times
10^{-3}$,  there  should  be   observable  effects    in   foreseeable
experiments    sensitive  to       $B(\mu \to   e\gamma      )    \sim
10^{-13}$~\cite{MEG,comment1} and to   a $\mu$-$e$ conversion  rate in
$^{48}_{22}{\rm Ti}$ (normalized   to the $\mu$  capture rate)  to the
$10^{-16}$  level~\cite{MECO,comment2}.  A much   higher signal in the
latter observable would  indicate  the presence of  possible  sizeable
non-decoupling terms of  the form $(a^2b^2v^2)/m^2_N$, which  dominate
for    $a,b\stackrel{>}{{}_\sim} 0.5$~\cite{IP}.  This       different
kinematic dependence  of  the two observables on   the  product of the
Yukawa couplings $ab$ would enable one to get some idea about the size
of   the heavy neutrino  mass  scale  $m_N$.  On  the  other hand, the
possible existence of electroweak-scale heavy neutrinos $N_{2,3}$ with
appreciable  $e$-Yukawa  couplings  $a \stackrel{>}{{}_\sim}   3\times
10^{-3}$ could be directly tested   by  studying their production   at
$e^+e^-$ linear colliders~\cite{BG}. Although it would be difficult to
produce $N_1$ directly,  a characteristic  signature of $N_{2,3}$   is
that they will decay predominantly  to $e$ and  $\mu$ leptons, but not
to $\tau$'s.  Moreover, since  $N_{2,3}$ play an important  synergetic
role in  resonantly  enhancing $\delta^\tau_{N_1}$,  potentially large
CP-violating  effects in their  decays  will determine the theoretical
parameters further.  Obviously, further detailed studies are needed to
obtain     a conclusive  answer    to    the   question  of    whether
electroweak-scale  resonant  $\tau$-leptogenesis could, in  principle,
provide a laboratory testable solution  to the cosmological problem of
the baryon asymmetry in the Universe.


\begin{references}
\bibitem{Example} For the specific set of model parameters: 
$m_N = 0.5$~TeV, 
$(\Delta  M_S)_{11} = 1.5\times 10^{-7}$~TeV, 
$(\Delta M_S)_{12} = -3\times 10^{-10}$~TeV, 
$(\Delta     M_S)_{13} = -10^{-10}$~TeV,  
$(\Delta M_S)_{22}  = 2.025\times   10^{-9}$~TeV, 
$(\Delta   M_S)_{23} = (3.375 + i0.405)\times 10^{-9}$~TeV, 
$(\Delta   M_S)_{33}  =  2.7\times   10^{-9}$~TeV,
$\varepsilon_e = (1.6382 + i0.178073)\times 10^{-6}$,
$\varepsilon_\mu = (2.04431 + i0.222217)\times 10^{-7}$,
$\varepsilon_\tau = (-3.00995 +   i0.327183)\times 10^{-7}$,  
$a  = 1.34541 \times 10^{-2}$, 
$b = 5.11257 \times 10^{-3}$, 
$c = 10^{-6}$,
we find in the parameterization of~\cite{PDG} the light-neutrino masses:
$m_{\nu_1} = 0.0504$~eV, $m_{\nu_2} = 0.0512$~eV,
$m_{\nu_3} = 0$~eV, and the mixing angles:
$\sin^2 \theta_{12} = 0.36$, $\sin^2 \theta_{23} = 0.34$, 
$\sin^2 \theta_{13} = 0.047$.
\bibitem{deltaN} $\delta_{N_1}$ is the total leptonic asymmetry in the
$N_1$ decays: $N_1 \to \Phi\: L_{e,\mu ,\tau}$. For the
scenario~\cite{Example}, it is $\delta_{N_1} \sim \delta^\tau_{N_1}$.
\bibitem{comment1} This result should be compared with that obtained in
ordinary resonant leptogenesis  scenarios~\cite{PU}, where  $a,b$ have
to be  factor 10 smaller,  leading correspondingly to the $10^4$ times
smaller prediction, $B(\mu \to e\gamma) \sim 10^{-17}$.
\bibitem{comment2}  For $a,b  \ll  1$,  the predicted  values for  the
appropriately normalized  transition  rate of $\mu$-$e$  conversion in
nuclei and for  $B  (\mu \to eee)$  are  about 2 orders of   magnitude
smaller than the ones for $B(\mu \to e\gamma )$.
\end{references}
\end{document}